\documentclass[aps,prl,twocolumn]{revtex4-2}
\usepackage{color,amsthm,amsmath,amsfonts,graphicx,bm,epsfig,float,siunitx,multirow,xr,enumerate,epstopdf}
\usepackage[breaklinks=true,colorlinks=true,linkcolor=blue,urlcolor=blue,citecolor=blue]{hyperref}
\usepackage[T1]{fontenc}

\begin{document}

\title{Nuclear Spin Induced Transparency}
\author{He-bin Zhang$^{1}$}
\author{Yuanjiang Tang$^{1}$}
\author{Yong-Chun Liu$^{1, 2}$}
\email{ycliu@tsinghua.edu.cn}
\affiliation{$^{1}$State Key Laboratory of Low-Dimensional Quantum Physics, Department of
Physics, Tsinghua University, Beijing 100084, P. R. China}
\affiliation{$^{2}$Frontier Science Center for Quantum Information, Beijing 100084, P.R.
China}
\date{\today }

\begin{abstract}
Electromagnetically induced transparency (EIT) is an important quantum
optical phenomenon which provides a crucial tool for light manipulation.
However, typically the transparency window is broad, limited by the
coherence time of the metastable state. Here we show that extremely narrow
transparency window can be realized using nuclear spin induced transparency
(NSIT), which is achieved by combining optical field, magnetic field and the
spin-exchange interaction between noble-gas nuclear spins and alkali-metal electronic spins. The
width of the NSIT window can be several orders of magnitude smaller than
that of conventional EIT, and even reaches sub-mHz range due to the long
coherence time of nuclear spins. The scheme holds great potential for
applications in slow light and magnetic field sensing.
\end{abstract}

\maketitle

Electromagnetically induced transparency (EIT) arises from the quantum
interference between transition paths driven by light fields in three-level
atoms or atom-like systems. Since the discovery of EIT~\cite%
{Harris1990_Nonlinear, Boller1991_Observation,
Harris1997_Electromagnetically}, various related applications have been
proposed including slow light~\cite{Hau1999_Light, Kash1999_Ultraslow,
Wu2004_Ultraslow, Harris1992_Dispersive, Lukin2001_Controlling,
Kasapi1995_Electromagnetically}, optical storage~\cite{Phillips2001_Storage,
Liu2001_Observation}, nonlinear effects~\cite{Schmidt1996_Giant,
Kash1999_Ultraslow}, and thus EIT has become an important research area in
quantum optics and quantum information \cite%
{Fleischhauer2005_Electromagnetically}. Besides, various EIT-like effects
based on different platforms have also been demonstrated \cite%
{Liu2017_Electromagnetically}, e.g., all-optical induced transparency~\cite%
{Smith2004_Coupled-resonator-induced, Yanik2004_Stopping, Totsuka2007_Slow,
Xu2006_Experimental, Xu2007_Breaking, Yang2009_All-Optical,
Xiao2009_Electromagnetically, Liu2017_Electromagnetically}, optomechanical
induced transparency \cite{Agarwal2010_Electromagnetically,
Weis2010_Optomechanically, Teufel2011_Circuit, Chang2011_Slowing,
Safavi-Naeini2011_Electromagnetically, Karuza2013_Optomechanically}, and plasmon
induced transparency~\cite{Zhang2008_Plasmon-Induced, Liu2009_Plasmonic}.
Nevertheless, the EIT window is typically broad due to the finite lifetime
of the metastable state, which limits its applications.

Meanwhile, the rare isotope of noble-gas atom, such as $^{3}$He and${}^{129}$%
Xe, has nonzero nuclear spin, which is isolated from the environment due to
the complete electronic shell and thus can have a coherence time longer than
one hour at or over room temperature~\cite{Gentile2017_Optically,
Gemmel2010_Ultra-sensitive, Heil2013_Spin}. 
Due to the unique properties, the research and applications on noble-gas nuclear spins have been extensively developed in various fields~\cite{Gentile2017_Optically}, including precision measurement~\cite{Kornack2005_Nuclear, Walker2016_Spin-Exchange-Pumped, Heil2017_Helium, Jiang2022_Floquet}, medical imaging~\cite{Albert1994_Biological, Middleton1995_MR, Couch2015_Hyperpolarized}, searches for new physics beyond the standard model~\cite{Lee2018_Improved, Safronova2018_Search, Chupp2019_Electric, Jiang2021_Search}, and quantum information technology~\cite{Katz2022_Quantum, Katz2022_Optical}.  
Recently, the coherent spin-exchange interaction between nuclear spins of noble-gas ensemble and electronic spins of alkali-metal ensemble is confirmed~\cite{Katz2022_Quantum, Katz2021_Coupling, Shaham2021_Strong}.
Based on this interaction, the theoretical schemes of long-lived entanglement generation ~\cite{Katz2020_Long-Lived} and quantum memory~\cite{Katz2022_Optical} have also been proposed.

However, since the nuclear spins are optically inaccessible, it is difficult
to construct conventional three-level EIT system with long-lived metastable
state directly. Here we demonstrate an analog of EIT by combining optical
field, magnetic field and the spin-exchange interaction between noble-gas nuclear spins and alkali-metal electronic spins, which we call nuclear spin induced
transparency (NSIT). This effect can be described by an effective four-level
model, through which we obtain analytical results. Very narrow transparency
window can be obtained, approaching the relaxation rate of noble-gas nuclear spins.
The center position of window is linearly dependent on the external magnetic
field strength. The corresponding dispersion curve is extremely steep,
resulting in strong slow light effect.

\begin{figure}[tbp]
\includegraphics[draft=false, width=1.0\columnwidth]{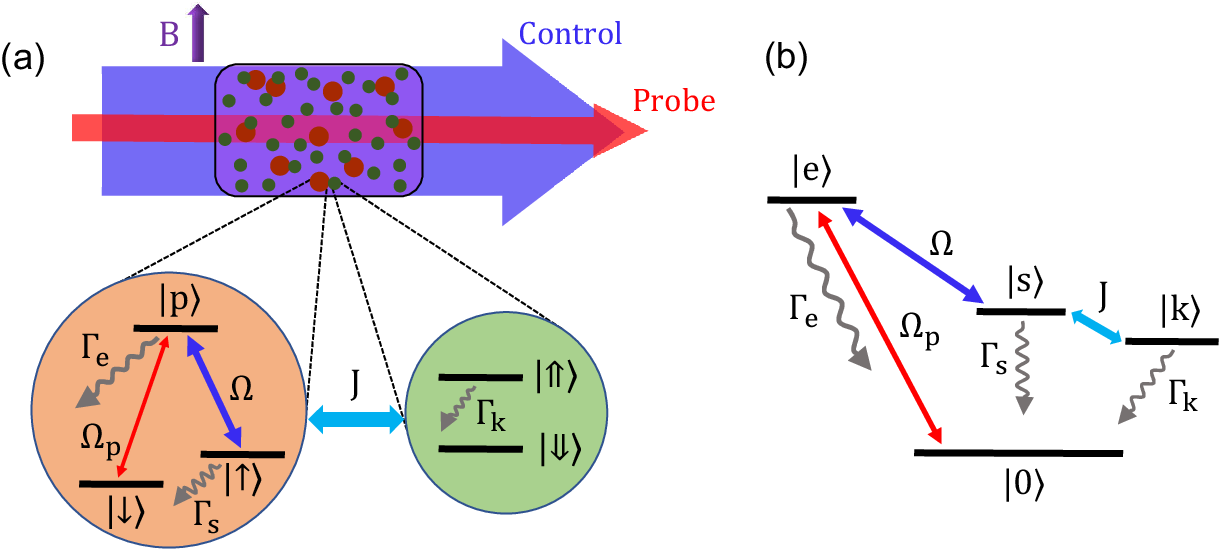}
\caption{(Color online) (a) Schematic diagram of the alkali-metal and
noble-gas mixture. The alkali metal atom is modeled as a $\Lambda $ system
consisting of a ground state $\left\vert \downarrow \right\rangle $, a
metastable state $\left\vert \uparrow \right\rangle $ and an excited state $%
\left\vert p\right\rangle $. The transitions $\left\vert \uparrow
\right\rangle \leftrightarrow \left\vert p\right\rangle $ and $\left\vert
\downarrow \right\rangle \leftrightarrow \left\vert p\right\rangle $ are
driven by a control field ($\Omega $) and a probe field ($\Omega _{p}$),
respectively. The noble-gas atom consists of down and up spin states $%
\left\vert \Downarrow \right\rangle $, $\left\vert \Uparrow \right\rangle $.
The coherent spin-exchange interactions between the noble-gas nuclear spins and the alkali-metal electronic spins are marked with skyey arrows ($J$). (b) Equivalent
energy levels of the hybrid system in (a), with $\left\vert 0\right\rangle
\equiv \left\vert \downarrow \right\rangle \left\vert \Downarrow
\right\rangle $, $\left\vert e\right\rangle \equiv \left\vert p\right\rangle
\left\vert \Downarrow \right\rangle $, $\left\vert s\right\rangle \equiv
\left\vert \uparrow \right\rangle \left\vert \Downarrow \right\rangle $, and
$\left\vert k\right\rangle \equiv \left\vert \downarrow \right\rangle
\left\vert \Uparrow \right\rangle $. The dissipation rates of the three
excited states satisfy ${\protect\Gamma _{e}}\gg {\protect\Gamma _{s}}\gg {%
	\protect\Gamma _{k}}$. }
\label{fig1}
\end{figure}

The system we consider is an alkali-metal vapor mixed with noble gas as
shown in Fig.~\ref{fig1}(a), with the atomic densities $n_a$ and $n_b$, respectively. The alkali-metal atom is modeled as a $\Lambda $
system consisting of a ground state $\left\vert \downarrow \right\rangle $, a
metastable state $\left\vert \uparrow \right\rangle $ and an excited state $%
\left\vert p\right\rangle $. A control field with frequency $\omega _{c}$
and a probe field with frequency $\omega _{p}$ drive the transitions $%
\left\vert \uparrow \right\rangle \leftrightarrow \left\vert p\right\rangle $
and $\left\vert \downarrow \right\rangle \leftrightarrow \left\vert
p\right\rangle $, respectively. 
The probe field is much weaker than the control field and propagates in the same direction as the control field to eliminate Doppler broadening.
The noble-gas atom with nuclear spin $1/2$
consists of down and up spin states $\left\vert \Downarrow \right\rangle $, $%
\left\vert \Uparrow \right\rangle $. An external magnetic field $B$ along
the quantization axis causes the Zeeman splitting of the alkali-metal and
noble-gas spins $\gamma_{s}B$ and $\gamma_{k}B$, respectively, where the gyromagnetic ratios satisfy $\gamma_{s}\gg \gamma_{k}$.
In this hybrid system, a coherent interaction has been
demonstrated between the collective alkali and noble-gas spins polarized along the same direction~\cite{Katz2022_Quantum,
Katz2021_Coupling, Shaham2021_Strong}. This interaction can be expressed as 
\begin{eqnarray}
{H_{\mathrm{sec}}}=&& \hbar \int {{d^3}\textbf{r} }  (\zeta\frac{[I]}{4} \left( {{X_{\downarrow \downarrow }}+{q_{I}}{X _{\uparrow \uparrow }}}\right) \left( {{X_{\Downarrow
	\Downarrow }}-{X_{\Uparrow \Uparrow }}}\right)  \notag \\
 &&+\zeta\frac{\sqrt{[I]}}{2} \left( {{X_{\downarrow \uparrow }}{X_{\Uparrow \Downarrow }}+H.c.}\right) ) , 
\label{eq-1a}
\end{eqnarray}
where $\textbf{r}_a$ and $\textbf{r}_b$ denote the positions of the alkali-metal and noble-gas atoms, respectively, and continuous atomic operators are introduced by ${X_{{\rm{mn}}}}(\textbf{r},t) = \sum\limits_{i = 1} {{{\left| m \right\rangle }_{\rm{i}}}{{\left\langle n \right|}_{\rm{i}}}} {\rm{ }}\delta (\textbf{r} - {\textbf{r}_i(t)})$ with $m,n \in \{  \downarrow , \uparrow ,{\rm{p}}, \Downarrow , \Uparrow \}$.
The first term leads to frequency shifts of atomic ensembles, and the second
term denotes the coherent spin-exchange interaction between alkali-metal and
noble-gas spins~\cite{Katz2022_Quantum, Katz2022_Optical}.

The dynamics of the system can be described by the Heisenberg-Langevin equation (See the
Supplemental Material \cite{[See the ][{ for details.}] NSIT_Supplemental}, Section I). In the weak excitation regime of the laser fields, one can obtain the motion equations for atomic excitations as 
\begin{eqnarray}
&&{\partial _t}{{\rm{\bar X}}_{ \downarrow {\rm{p}}}} =  - (i{\Delta _{\rm{e}}} + \frac{{{\Gamma _{\rm{e}}}}}{2}){{\rm{\bar X}}_{ \downarrow {\rm{p}}}} - i{\Omega _{\rm{p}}}{{\rm{\bar X}}_{ \downarrow  \downarrow }} - i\Omega {{\rm{\bar X}}_{ \downarrow  \uparrow }}  + {\bar F_{ \downarrow {\rm{p}}}} , \notag \\
&&{\partial _t}{{\rm{\bar X}}_{ \downarrow  \uparrow }} =  - (i{\Delta _{\rm{s}}} + \frac{{{\Gamma _{\rm{s}}}}}{2}){{\rm{\bar X}}_{ \downarrow  \uparrow }} - i{\Omega ^*}{{\rm{\bar X}}_{ \downarrow p}} - i J {{\rm{\bar X}}_{ \Downarrow  \Uparrow }} + {\bar F_{ \downarrow  \uparrow }}\notag \\
&&{\partial _t}{{\rm{\bar X}}_{ \Downarrow  \Uparrow }} =  - (i{\Delta _{\rm{k}}} + \frac{{{\Gamma _{\rm{k}}}}}{2}){{\rm{\bar X}}_{ \Downarrow  \Uparrow }} - i J {{\rm{\bar X}}_{ \downarrow  \uparrow }} + {\bar F_{ \Downarrow  \Uparrow }}.
\label{eq-1}
\end{eqnarray}
with ${{\rm{\bar X}}_{ \downarrow \downarrow}} = {{\rm{X}}_{ \downarrow \downarrow}}/\sqrt {{p_a}{n_a}}$, ${{\rm{\bar X}}_{ \downarrow {\rm{p}}}} = {{\rm{X}}_{ \downarrow {\rm{p}}}}/\sqrt {{p_a}{n_a}}$, ${\rm{ }}{{\rm{\bar X}}_{ \downarrow  \uparrow }} = {{\rm{X}}_{ \downarrow  \uparrow }}/\sqrt {{p_a}{n_a}}$, and ${{\rm{\bar X}}_{ \Downarrow  \Uparrow }} = {{\rm{X}}_{ \Downarrow  \Uparrow }}/\sqrt {{p_b}{n_b}}$.  
$\Omega $ and $\Omega _{p}$ denote the Rabi frequencies of the control
and probe fields, respectively, and $J=\zeta \sqrt{{p_a p_b}{n_a n_b}/4}$ denotes the spin-exchange rate.
The parameters ${\Gamma _{e,s,k}}$ denotes the dissipation rates. 
The laser detunings are defined as ${\Delta _{\rm{e}}} = {\tilde\omega_e} - {{\rm{\omega }}_{\rm{p}}}$,  ${\Delta _{\rm{s}}} = {\tilde\omega_s} - ({{\rm{\omega }}_{\rm{p}}} - {{\rm{\omega }}_{\rm{c}}})$, and ${\Delta _{\rm{k}}} = {\tilde\omega_k} - ({{\rm{\omega }}_{\rm{p}}} - {{\rm{\omega }}_{\rm{c}}})$, where $\tilde\omega_e$,  $\tilde\omega_s$, and $\tilde\omega_k$ denote the overall frequencies of the energy levels, containing the frequency shifts due to the magnetic field and the spin-exchange interaction.  
The overall frequency difference between the alkali and noble-gas spins can be compensated by a specific magnetic field $B=B_{0}$, resulting in $\tilde\omega_s=\tilde\omega_k$, i.e., $\Delta _{s}=\Delta _{k}$. 
Furthermore, for the convenience of the theoretical analysis, we assume that the detunings satisfy a simple relationship $\Delta _{e}=\Delta _{s}=\Delta _{k}$ at $B=B_{0}$, which can be realized by adjusting the frequency of the control field. Accordingly, we deduce that ${\Delta _{s}}={\Delta _{e}}+{\gamma_{s}}{\widetilde{B}}$, ${\Delta _{k}}={\Delta _{e}}+{\gamma_{k}}{\widetilde{B}}$ hold for a general magnetic field $B=B_{0}+\widetilde{B}$ and tunable probe frequency $\omega_p$.
The last term of each equation represents the quantum noise. 

By solving Eq.~(\ref{eq-1}), we obtain that the normalized complex
susceptibility $\chi ={\chi _{1}}+{\chi _{2}}+{\chi _{3}}$ of the alkali atoms, which is proportional to the expected value of $\bar X_{ \downarrow {\rm{p}}}$, consists of three
components (See the Supplemental Material \cite{NSIT_Supplemental}, Section
II):
\begin{eqnarray}
&&\chi _{1}=-i\eta \frac{{1}}{{i\left( {{\Delta _{e}}-\frac{{4\gamma_{s}\widetilde{B}{{\left\vert \Omega \right\vert }^{2}}}}{{{\Gamma _{e}}^{2}}}}\right) +%
	\frac{{\widetilde{\Gamma} _{e}}}{2}}},  \notag \\
&&\chi _{2}=-\eta \frac{{8\left( {2\gamma_{s}\widetilde{B}-i\frac{{\Gamma _{e}}}{2}}%
	\right) {{\left\vert \Omega \right\vert }^{2}}}}{{\left( {i\left( {{\Delta
				_{e}}+\gamma_{s}\widetilde{B}}\right) +\frac{{\widetilde{\Gamma} _{s}}}{2}}\right) {\Gamma _{e}}%
	^{3}}},  \label{eq-3} \\
&&\chi _{3}=-i\eta \frac{{4J^{2}{{\left\vert \Omega \right\vert }^{2}}}}{{%
	\left( {i\left( {{\Delta _{e}}-\bar{\omega} }\right) +\frac{{\widetilde{\Gamma} _{k}}}{2}}%
	\right) {{\left( {i\gamma_{s}\widetilde{B}{\Gamma _{e}}+\frac{1}{2}{\Gamma _{e}}{\Gamma _{s}}+2{{\left\vert \Omega \right\vert }^{2}}}\right) }^{2}}}}.
\notag
\end{eqnarray}%
Here $\chi _{1}$ and $\chi _{2}$ correspond to the susceptibility of EIT
with the background widths ${\widetilde{\Gamma} _{e}}={\Gamma _{e}}-4{\left\vert \Omega \right\vert }^{2}/\Gamma_e $ 
and the EIT window width ${\widetilde{\Gamma} _{s}}={\Gamma _{s}}+4{\left\vert \Omega \right\vert }^{2}/\Gamma_e $. The last term $\chi _{3}$ represents the nuclear spin
induced absorption and transparency effects, which originate from the
spin-exchange interaction, and manifest the action of noble-gas nuclear spin on the
optical transition of the alkali-metal atom. The corresponding frequency
shift $\bar{\omega} $ and the width $\widetilde{\Gamma} _{k}$ are given by
\begin{eqnarray}
&&\bar{\omega} =-{\gamma_{k}}\widetilde{B}+{\gamma_{s}}\widetilde{B}\frac{{{J}^{2}{\Gamma _{e}}%
	^{2}}}{{{\ \gamma_{s}^{2}\widetilde{B}^{2}}{\Gamma _{e}}^{2}+{{({\frac{1}{2}{\Gamma
					_{e}}{\Gamma _{s}}+2{{\left\vert \Omega \right\vert }^{2}}})}^{2}}}},  \notag
\\
&&{\widetilde{\Gamma} _{k}}={\Gamma _{k}}+\frac{{{J}^{2}{\Gamma _{e}}({{\Gamma _{e}}{%
			\Gamma _{s}}+4{{\left\vert \Omega \right\vert }^{2}}})}}{{{\gamma_{s}^{2}\widetilde{B}^{2}}{\Gamma _{e}}^{2}+{{({\frac{1}{2}{\Gamma _{e}}{\Gamma _{s}}+2{{%
						\left\vert \Omega \right\vert }^{2}}})}^{2}}}}.  \label{eq-4}
\end{eqnarray}

\begin{figure}[tbp]
\includegraphics[draft=false, width=1\columnwidth]{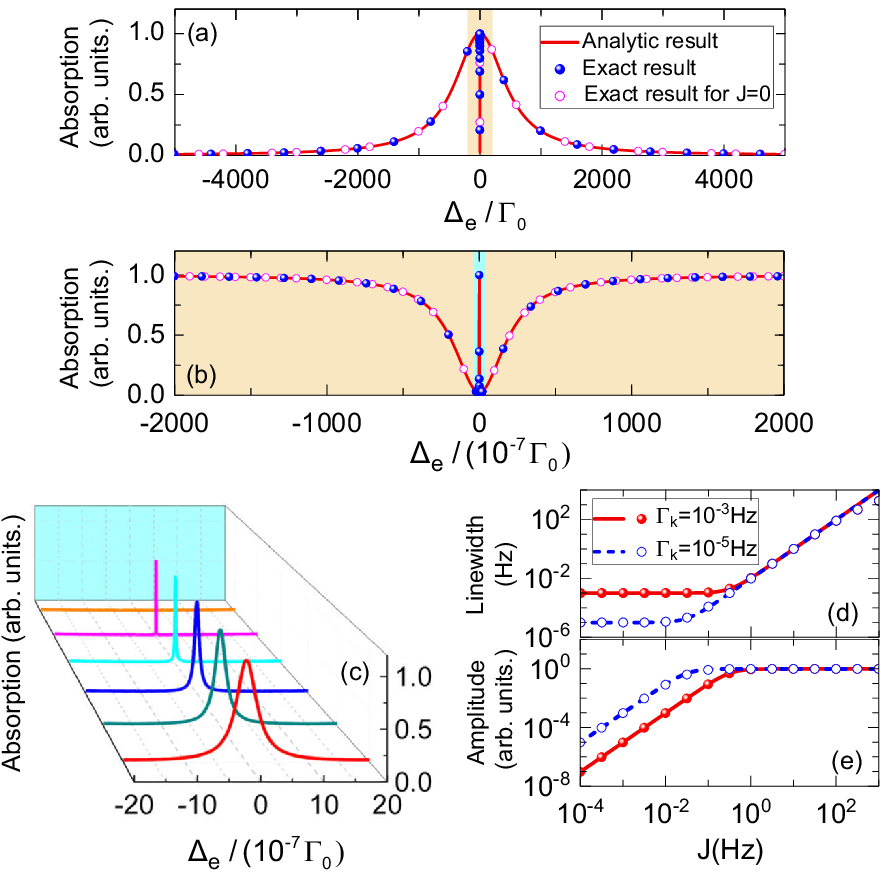}
\caption{(Color online) (a) Normalized absorption spectrum of the optical
	transition $\left\vert \downarrow \right\rangle \leftrightarrow \left\vert p\right\rangle $ at $\protect\widetilde{B}=0$, i.e., ${\Delta _{s}}={\Delta_{k}}$ $={\Delta _{e}}$. Other parameters satisfy ${\protect\Gamma_{e}}=10^{3}\Gamma_0,{\protect\Gamma _{s}}=10^{-6}\Gamma_0,{\protect\Gamma _{k}}=10^{-10}\Gamma_0, \Omega =10^{-1}\Gamma_0, J=10^{-6}\Gamma_0$, and $\gamma_{s}=100\gamma_{k}$, with $\Gamma_0$ denoting the natural linewidth of the alkali-metal atom. 
	(b) Zoom-in view of $10^{-6}$ region with a brown background in (a), where the absorption peak of NSIA is revealed in the center cyan region. 
	(c) Normalized absorption spectrum for $J=20, 15, 10, 5, 1, 0\times10^{-7}\Gamma_0$ from front to back. 
	(d) Linewidth and (e) amplitude of NSIA signal as functions of $J$. The lines and circles in (d) and (e) correspond to the analytical and numerical results, respectively. Natural linewidth $\Gamma_0$ is valued as $\Gamma_0=10^{7}\textrm{Hz}$ according to typical alkali-metal atoms to facilitate the estimation of the specific features of NSIA. Other parameters in (c)-(e) are the same as that in (a). }
\label{fig2}
\end{figure}

In Fig.~\ref{fig2}, we focus on the nuclear spin induced absorption (NSIA)
effect at the resonance condition with $B=B_{0}$. As plotted in Fig.~\ref%
{fig2}(a), there appears a dip at the center of the absorption spectrum,
which corresponds to the conventional EIT window. When zooming in on the
spectrum near the EIT dip [Fig.~\ref{fig2}(b)], we can find that a very
narrow absorption peak appears at the center of the EIT dip. By comparison with the case of $J=0$, we see that this peak corresponds to the contribution of $\chi _{3}$ in Eq.~(\ref{eq-3}) and thus originates from the coherent spin-exchange interaction between the alkali-metal and noble-gas spins. Specifically, the alkali-metal atom exhibits a normal EIT phenomenon due to the quantum interference of the transition amplitude for two different pathways.
Further, at the resonance regime of the spin-exchange interaction, the optical transition in the alkali-metal atom strongly couples with the transition in the nuclear spin of noble-gas atom, which leads to a different quantum interference and thereby NSIA.

The performance of the NSIA signal is closely related to the exchange rate $J$. The
spectra of the NSIA signals for various $J$ are plotted in Fig.~\ref{fig2}(c),
and the corresponding linewidth and amplitude as functions
of the exchange rate $J$ are shown in Figs.~\ref{fig2}(d) and (e),
respectively. It reveals that when $J$ is small, the linewidth of the
absorption peak can approach the lower limit, i.e., the decoherence rate of
noble-gas nuclear spin $\Gamma _{k}$, despite of the decrease of the amplitude. As $J
$ increases, the linewidth of the absorption peak increases, and meanwhile
the amplitude can reach and maintain the maximum.

\begin{figure}[tbp]
\includegraphics[draft=false, width=1\columnwidth]{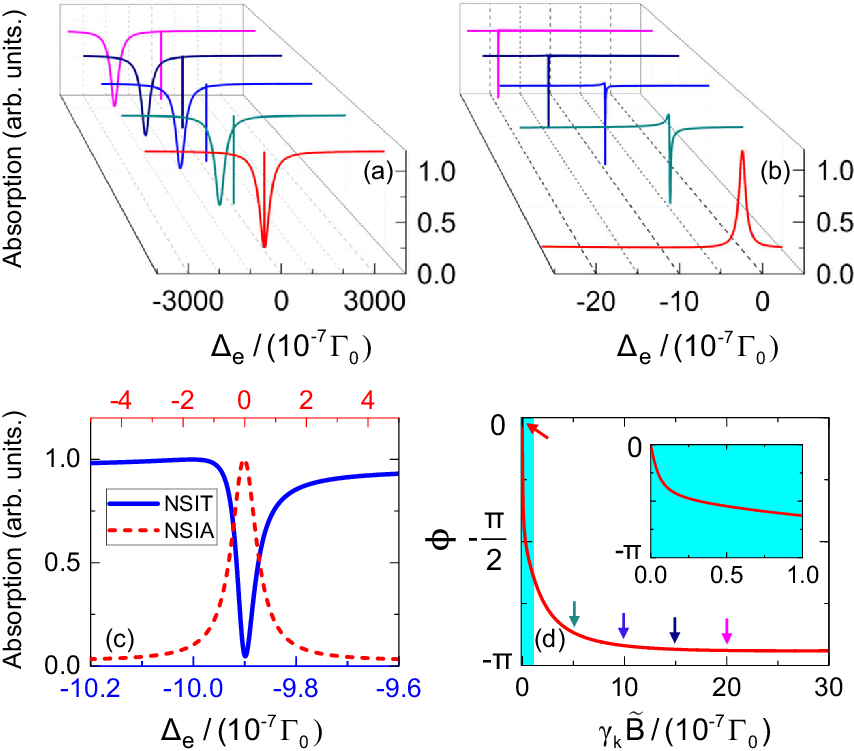}
\caption{(Color online) (a) Normalized absorption spectrum of the optical
transition $\left| \downarrow \right\rangle \leftrightarrow \left| p
\right\rangle$ for $\protect\widetilde{B} =0,5,10,15,20\times10^{-7}\Gamma_0/\gamma_k$ from front to back. 
(b) Zoom-in view of the region where the resonance signals induced by the spin-exchange interaction in (a) are involved. 
(c) The blue solid line and the red dashed line correspond to NSIT for $\protect\widetilde{B}=10^{-6}\Gamma_0/\gamma_k$ and NSIA for $\protect\widetilde{B} =0$ in (b), respectively. (d)
Relative phase $\protect\phi$ between different transition channels at $\Delta_e=\protect\bar{\omega}$ as a function of the Zeeman splitting $\gamma_k \protect%
\widetilde{B} $. The arrows indicate the positions corresponding to the absorption curves in (a) and (b). The inset shows the zoom-in view of the region with a cyan background. Other parameters are the same as Fig.~%
\protect\ref{fig2}(a). }
\label{fig3}
\end{figure}

Next we focus on the case without the resonance condition, for an arbitrary
magnetic field $B=B_{0}+\widetilde{B}$. In Fig.~\ref{fig3}(a) and (b), we plot
the absorption spectrum corresponding to different magnetic field strengths.
It reveals that in addition to the shift of EIT window, the position of the
absorption peak of NSIA also shifts with the change of the magnetic field.
According to Eq.~(\ref{eq-4}), the shift $\bar{\omega} $ of the NSIA peak consists
of two terms. The first term corresponds to the Zeeman splitting of the
noble-gas spin induced by $\widetilde{B}$, and the second term is a nonlinear effect associated
with the Zeeman splitting of the alkali-metal spin. Under practical experimental conditions, the exchange rate $J$ can be significantly weaker than the power broadening ${\left\vert \Omega \right\vert }^{2}/\Gamma_e$ of the control field, and then the nonlinear effect is negligible in the weak magnetic field region. Accordingly, the shift of the narrow peak is mainly determined by the Zeeman splitting of the noble-gas spin, which is linearly dependent on the external magnetic field strength (see Fig.~\ref{fig3}(b)).  Meanwhile, the shape of the NSIA signal changes with the magnetic field. For relatively
large $\widetilde{B}$, this narrow absorption peak transforms into a dip, as
shown in Fig.~\ref{fig3}(b). Besides, the minimum absorption is extremely close to zero
as plotted in Fig.~\ref{fig3}(c), which indicates that the significant NSIT effect
occurs. 
Since the decoherence rates of the alkali and noble-gas
spins satisfy $\Gamma _{k}\ll \Gamma _{s}$, the width of NSIT and NSIA can
be several orders of magnitude smaller than the width of the conventional EIT
window. Thereby, by measuring these signals induced by nuclear spin, the magnetic field detection with unprecedented accuracy can be realized. 

To understand the quantum interference in NSIA and NSIT, we derive the
probability amplitudes for state $\left\vert e\right\rangle $ according to
Eq.~(\ref{eq-1}), from which we obtain $\left\langle{\bar X}_{ \downarrow p} \right\rangle=\frac{{i{\Omega _{p}}%
\left( {\Gamma _{s}/2+i{\Delta _{s}}}\right) \left\langle{\bar X}_{ \downarrow \downarrow} \right\rangle+J\Omega
\left\langle{{\rm{\bar X}}_{ \Downarrow  \Uparrow }} \right\rangle}}{{\left( {\Gamma _{e}/2+i{\Delta _{e}}}\right) \left( {%
	\Gamma _{s}/2+i{\Delta _{s}}}\right) +{{\left\vert \Omega \right\vert }^{2}}}%
}$, which consists of two terms and indicates that the excitation of the
alkali-metal atom results from two transition channels. The first term
represents the contribution of the optical transition $\left\vert \downarrow
\right\rangle \leftrightarrow \left\vert p\right\rangle $ driven by the
probe field, and the second term represents the contribution of the
noble-gas nuclear spin excitation through spin-exchange interaction and the
subsequent transition driven by the control field. The relative phase $\phi $
between these two coherent transition channels at $\Delta _{e}=\bar{\omega} $,
derived as $\phi =\arg [\frac{J\Omega \left\langle{{\rm{\bar X}}_{ \Downarrow  \Uparrow }} \right\rangle}{i{\Omega _{p}}\left( {%
\Gamma _{s}/2+i{\Delta _{s}}}\right) \left\langle{\bar X}_{ \downarrow \downarrow} \right\rangle}]$, is plotted in Fig.~%
\ref{fig3}(d). It reveals that when $\widetilde{B}=0$, the constructive
interference occurs between these two transition channels, which is
consistent with the appearance of NSIA. Meanwhile, as $\widetilde{B}$
increases, the relative phase $\phi $ increases. When the magnetic field is
large enough that the signal caused by the spin-exchange interaction can be
distinguished from the EIT window, the destructive interference between the
two coherent transition channels occurs, which explains the appearance of
NSIT. 

\begin{figure}[tbp]
\includegraphics[draft=false, width=1\columnwidth]{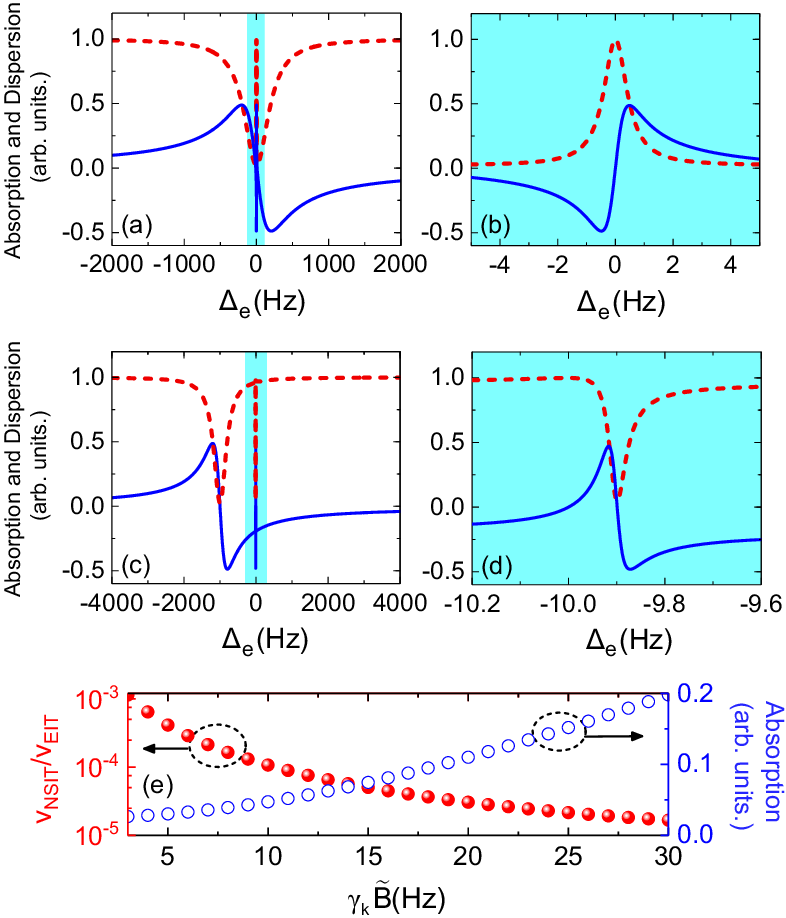}
\caption{(Color online) Normalized absorption-dispersion spectrums for NSIA
at $\protect\widetilde{B} =0$ in (a)-(b), and NSIT at $\protect\widetilde{B} = 10^{-6}\Gamma_0/\gamma_k$ in (c)-(d) with $\Gamma_0=10^{7}\textrm{Hz}$, where the blue solid line and red dashed line
represent dispersion and absorption, respectively. (b) and (d) are the zoom-in views of about $1/20$ and $10^{-3}$ regions with a cyan background in (a) and (c), respectively, where NSIA and NSIT are revealed. (e) Minimum absorption at NSIT window (blue open circles), and the ratio
of the light group velocities at NSIT and EIT windows (red solid circles) as
functions of the Zeeman splitting $\gamma_k \protect\widetilde{B} $. Other parameters
are the same as Fig.~\protect\ref{fig2}(a). }
\label{fig4}
\end{figure}

The absorption-dispersion spectrums corresponding to NSIA and NSIT are exhibited in Fig.~\ref{fig4},
where the values of the main physical parameters adopted are referenced to actual systems~\cite{Katz2021_Coupling, Shaham2021_Strong}.
In Figs.~\ref{fig4}(a) and (b), we show the case where the external magnetic field
satisfies $B=B_{0}$, and thus NSIA arises as mentioned above. It reveals
that in the region of NSIA, despite the difference in sign, the slope of the
dispersion curve is much larger than that at the EIT window in Figs.~\ref{fig4}(a). In Figs.~\ref%
{fig4}(c) and (d), we show the case where the difference between the
frequency shifts of the alkali and noble-gas spins are significant, so that
NSIT occurs. It reveals that besides the same sign, the slope of the
dispersion curve at the NSIT window is several orders of magnitude greater
than that at the EIT window.

The steep dispersion curve leads to a small group velocity of the light pulse,
corresponding to strong slow light effect, which has important applications
in various fields including \ optical delay~\cite{Tucker2005_Slow-light} and
nonlinear effects~\cite{Schmidt1996_Giant, Kash1999_Ultraslow, Baba2008_Slow}%
. In Fig.~\ref{fig4}(e) we plot the ratio of light group velocities at the
NSIT and EIT windows, and the corresponding minimum absorption at NSIT
window. It reveals that the light group velocity at the NSIT window can be
many orders of magnitude smaller than that at EIT window, which results from
the fact that the decoherence rate of the noble-gas spin is much smaller
than that of the alkali spin. Besides, as the magnetic field increases, the
dispersion curve becomes sharper, resulting in a more pronounced slow light
effect.

Different combinations of alkali and noble-gas atoms can be chosen as
candidates for protocol implementation, and the experimental configuration for observing EIT in atomic gas can act as a reference for the observation of NSIT and related effects. 
The appearance of NSIT and NSIA depends on the coherent spin-exchange interaction between the noble-gas nuclear spins and alkali-metal electronic spins, which has been
confirmed experimentally using a K-He mixed vapor cell~\cite{Shaham2021_Strong}. Further, the bi-directional interface between light and noble-gas nuclear spins has been identified, and the optical signal originating from noble-gas nuclear spin has been observed in the transmission spectrum of signal light~\cite{Katz2021_Coupling}. We demonstrate that NSIT phenomenon can arise under relaxed
conditions, and the signal widths are tunable from sub-mHz to tens of Hz or moreover (See the
Supplemental Material \cite{NSIT_Supplemental}, Section III), which further
facilitate the corresponding proof-of-principle experiment and potential
applications.
Doppler effect due to atomic motions can be suppressed based on the Doppler-free technique, which is commonly applied when observing EIT in atomic gas~\cite{Li1995_Observation, Gea-Banacloche1995_Electromagnetically}. Based on this technique, the narrow NSIT signal can be well maintained (See the Supplemental Material \cite{NSIT_Supplemental}, Section IV).  In addition, we discuss the role of imperfect polarization on NSIT phenomenon (See the Supplemental Material \cite{NSIT_Supplemental}, Section V), and demonstrate the stability of NSIT.
Therefore, the proposed scheme exhibits practical feasibility under the current experimental conditions. 

In summary, we demonstrate that the NSIT phenomenon can occur in the mixture
of alkali-metal and noble-gas atoms, which originates from the effect of the
nuclear spin of noble-gas atoms on the optical transition of alkali-metal
atoms through the spin-exchange interaction between the alkali-metal and
noble-gas spins. We construct an effective four-level model to describe the
NSIT phenomenon and obtain analytical results. The width of the NSIT window
can be several orders of magnitude smaller than that of conventional EIT,
and even approaches the relaxation rate of noble-gas nuclear spin on the order of
sub-mHz range~\cite{Gentile2017_Optically, Walker1997_Spin-exchange}. The center position of the window is linearly dependent on
the external magnetic field strength, making it a good candidate for
high-sensitivity magnetic field sensing. Besides, the dispersion curve
corresponding to the transparent window is extremely sharp, which can lead
to a slow light effect where the group velocity of the light pulse can be
several orders of magnitude smaller than that in conventional EIT. The NSIT
mechanism provides an important method for various applications including
long-time light storage and high-performance sensing.

\begin{acknowledgments}
This work is supported by the National Key R\&D Program of China (Grant No. 2023YFA1407600), and the National Natural Science Foundation of China (NSFC) (Grants No. 12275145, No. 92050110, No. 91736106, No. 11674390, and No. 91836302).
\end{acknowledgments}

\bibliography{bib_NSIT}

\end{document}




\title{Supplemental Material for ``Nuclear Spin Induced Transparency"}

\author{He-bin Zhang$^{1}$}
\author{Yuanjiang Tang$^{1}$}
\author{Yong-Chun Liu$^{1, 2}$}
\email{ycliu@tsinghua.edu.cn}
\affiliation{$^{1}$State Key Laboratory of Low-Dimensional Quantum Physics, Department of Physics, Tsinghua University, Beijing 100084, P. R. China}
\affiliation{$^{2}$Frontier Science Center for Quantum Information, Beijing 100084, P.R. China}

\date{\today}

\begin{abstract}
This supplementary material contains five parts: 
\uppercase\expandafter{\romannumeral1}. Hamiltonian of the atomic ensembles and the derivation of the motion equations adopted in the main text are introduced. \uppercase\expandafter{\romannumeral2}. By Laplace transform, we simplify the analytical expression of the complex susceptibility, which can facilitate the verification and understanding of the properties of the resonance signal induced by the coherent spin-exchange interaction between the alkali and noble-gas spins. 
\uppercase\expandafter{\romannumeral3}. We investigate the absorption properties of the optical transition $\left|  \downarrow  \right\rangle  \leftrightarrow \left| p \right\rangle$ over a wide parameter range and analyze the physical parameters affecting the linewidth and amplitude of NSIT signal, which demonstrate that NSIT can arise under relaxed conditions and is highly tunable.   
\uppercase\expandafter{\romannumeral4}. We analyze the influence of Doppler effect on the NSIT.
\uppercase\expandafter{\romannumeral5}. We discuss the role of imperfect polarization on the NSIT.
\end{abstract}


\maketitle
\tableofcontents

\section{Motion equations}

The Hamiltonian describing the free energies of the system and the interactions between the atoms and the laser fields in the rotating frame of the control and probe fields is given by 
\begin{eqnarray}
{H_{\rm{e}}} = \hbar \int {{d^3}\textbf{r}({\Delta'_e}{{\rm{X}}_{{\rm{pp}}}} +  } {\Delta'_s}{{\rm{X}}_{ \uparrow  \uparrow }} + {\Delta'_k}{{\rm{X}}_{ \Uparrow  \Uparrow }} + (\Omega{{X}_{{\rm{p}} \uparrow }} + {\Omega _p}{{X}_{{\rm{p}} \downarrow }} + {\rm{H}}{\rm{.c}}{\rm{.}})).
\label{eq-S1a}
\end{eqnarray}
with the detunings ${{\Delta_e'}} = {\omega _e} - {\omega _p},{{\Delta_s '}} = {\gamma_s}B - ({\omega _p} - {\omega _c}), {{\Delta_k '}} = {\gamma_k}B - ({\omega _p} - {\omega _c})$. $\Omega $ and $\Omega _{p}$ denote the Rabi frequencies of the control
and probe fields, respectively. Continuous atomic operators of the atomic ensembles are introduced by 
\begin{eqnarray}
{X_{{\rm{mn}}}}(\textbf{r},t) = \sum\limits_{i = 1} {{{\left| m \right\rangle }_{\rm{i}}}{{\left\langle n \right|}_{\rm{i}}}} {\rm{ }}\delta (\textbf{r} - {\textbf{r}_i(t)}),
\label{eq-S1a2}
\end{eqnarray}
with $m,n \in \{  \downarrow , \uparrow ,{\rm{p}}, \Downarrow , \Uparrow \} $.
For polarized gases, a coherent spin-exchange interaction between the collective alkali and noble-gas atomic ensembles is demonstrated~\cite{Katz2022_Quantum, Katz2021_Coupling, Shaham2021_Strong}, and can be expressed as
\begin{eqnarray}
{H_{\mathrm{sec}}}=&& \hbar \int {{d^3}\textbf{r} }  (\zeta\frac{[I]}{4} \left( {{X_{\downarrow \downarrow }}+{q_{I}}{X _{\uparrow \uparrow }}}\right) \left( {{X_{\Downarrow
	\Downarrow }}-{X_{\Uparrow \Uparrow }}}\right) +\zeta\frac{\sqrt{[I]}}{2} \left( {{X_{\downarrow \uparrow }}{X_{\Uparrow \Downarrow }}+H.c.}\right) ) ,
\label{eq-1a}
\end{eqnarray}
where $q_I=([I]-2)/[I]$ with alkali-metal nuclear spin $I$ ($[I]\equiv 2I+1$). $\zeta$ denotes the microscopic spin-exchange strength.
The motion equation of the atomic system is given by the following Heisenberg-Langevin equation
\begin{eqnarray}
{\partial _t}{{\rm{X}}_{{\rm{mn}}}} = \frac{i}{\hbar }[H_e+H_{sec},{{\rm{X}}_{{\rm{mn}}}}] - \frac{{{\Gamma _{{\rm{mn}}}}}}{2}{{\rm{X}}_{{\rm{mn}}}} + {F_{{\rm{mn}}}}.
\label{eq-S1b}
\end{eqnarray}
The second term generally represented relaxation processes. The last term of each equation represents the quantum noise with the noise operator $F_{{\rm{mn}}}$. 
Accordingly, the explicit motion equations for the operators relating the optical excitation and spin polarization of the alkali-metal atoms are given by 
\begin{eqnarray}
&&{\partial _t}{{\rm{X}}_{ \downarrow {\rm{p}}}} =  - (\frac{\Gamma_e}{2} + i\Delta'_e - i\frac{{[I]}}{4}\zeta ({{\rm{X}}_{ \Downarrow  \Downarrow }} - {{\rm{X}}_{ \Uparrow  \Uparrow }})){{\rm{X}}_{ \downarrow {\rm{p}}}} - i{\Omega _{\rm{p}}}({{\rm{X}}_{ \downarrow  \downarrow }} - {{\rm{X}}_{{\rm{pp}}}}) - i\Omega {{\rm{X}}_{ \downarrow  \uparrow }} + i\frac{{\sqrt {[I]} }}{2}\zeta {{\rm{X}}_{ \uparrow {\rm{p}}}}{{\rm{X}}_{ \Downarrow  \Uparrow }} + {F_{ \downarrow {\rm{p}}}},   \notag \\
&&{\partial _t}{{\rm{X}}_{ \downarrow  \uparrow }} =  - (\frac{\Gamma_s}{2} + i{\Delta'_s} + i\frac{{{q_I} - 1}}{4}[I]\zeta ({{\rm{X}}_{ \Downarrow  \Downarrow }} - {{\rm{X}}_{ \Uparrow  \Uparrow }})){{\rm{X}}_{ \downarrow  \uparrow }} - i{\Omega ^*}{{\rm{X}}_{ \downarrow p}} - i\frac{{\sqrt {[I]} }}{2}\zeta ({{\rm{X}}_{ \downarrow  \downarrow }} - {{\rm{X}}_{ \uparrow  \uparrow }}){{\rm{X}}_{ \Downarrow  \Uparrow }} + i{\Omega _{\rm{p}}}{{\rm{X}}_{{\rm{p}} \uparrow }} + {F_{ \downarrow  \uparrow }},   \notag \\
&&{\partial _t}{{\rm{X}}_{ \Downarrow  \Uparrow }} =  - (\frac{\Gamma_k}{2} + i\Delta'_k - i\frac{{[I]}}{2}\zeta ({{\rm{X}}_{ \downarrow  \downarrow }} + {q_I}{{\rm{X}}_{ \uparrow  \uparrow }})){{\rm{X}}_{ \Downarrow  \Uparrow }} - i\frac{{\sqrt {[I]} }}{2}\zeta {{\rm{X}}_{ \downarrow  \uparrow }}({{\rm{X}}_{ \Downarrow  \Downarrow }} - {{\rm{X}}_{ \Uparrow  \Uparrow }}) + {F_{ \Downarrow  \Uparrow }},
\label{eq-S1c}
\end{eqnarray}
The parameters ${\Gamma _{e}}$, ${\Gamma _{s}}$, and ${\Gamma _{k}}$ denote the relaxation rates of alkali-metal excitations, alkali-metal spins, and noble-gas spins, respectively.
The addition of high-pressure noble gases can lead to a relaxation of alkali-metal excitations beyond GHz~\cite{Katz2021_Coupling, Shaham2021_Strong}, far exceeding the natural linewidth due to spontaneous decay.
Due to the isolation from the environment, the relaxation rate of noble-gas spins is much smaller than that of alkali-metal spins, and the coherence time can reach several hundreds of hours at or over room temperature~\cite{Gentile2017_Optically, Walker1997_Spin-exchange}.

Considering that the control field is in the weak excitation regime, i.e., $\Omega  \ll {\Gamma _{\rm{e}}}$, and that the strength of the probe field is negligible compared to the control field, one can assume that ${{\rm{X}}_{{\rm{pp}}}} \approx {{\rm{X}}_{{\rm{p}} \uparrow }} \approx 0$. 
We focus on the regime of highly polarized spin ensembles, so the equations can be simplified under the Holstein-Primakoff approximation~\cite{Hammerer2010_Quantum, Holstein1940_Field}. 
Accordingly, the collective operator of alkali spins can be represented classically as ${{\rm{X}}_{ \downarrow  \downarrow }} \approx {p_{\rm{a}}}{n_{\rm{a}}}{\rm{, }}{{\rm{X}}_{ \uparrow  \uparrow }} \approx (1 - {p_{\rm{a}}}){n_{\rm{a}}}$, where $n_a$ and $p_a$ denote the density of alkali atoms and the polarization degree of alkali spins, respectively. Similarly, the collective operator of noble-gas spins can be represented classically as ${{\rm{X}}_{ \Downarrow  \Downarrow }} \approx {p_{\rm{b}}}{n_{\rm{b}}}{\rm{, }}{{\rm{X}}_{ \Uparrow  \Uparrow }} \approx (1 - {p_{\rm{b}}}){n_{\rm{b}}}$, where $n_b$ and $p_b$  denote the density of noble-gas atoms and the polarization degree of noble-gas spins, respectively.
Consequently, the simplified equations can be obtained as  
\begin{eqnarray}
&&{\partial _t}{{\rm{\bar X}}_{ \downarrow {\rm{p}}}} =  - (i{\Delta _{\rm{e}}} + \frac{{{\Gamma _{\rm{e}}}}}{2}){{\rm{\bar X}}_{ \downarrow {\rm{p}}}} - i{\Omega _{\rm{p}}}{{\rm{\bar X}}_{ \downarrow  \downarrow }} - i\Omega {{\rm{\bar X}}_{ \downarrow  \uparrow }}  + {\bar F_{ \downarrow {\rm{p}}}} , \notag \\
&&{\partial _t}{{\rm{\bar X}}_{ \downarrow  \uparrow }} =  - (i{\Delta _{\rm{s}}} + \frac{{{\Gamma _{\rm{s}}}}}{2}){{\rm{\bar X}}_{ \downarrow  \uparrow }} - i{\Omega ^*}{{\rm{\bar X}}_{ \downarrow p}} - i J {{\rm{\bar X}}_{ \Downarrow  \Uparrow }} + {\bar F_{ \downarrow  \uparrow }}\notag \\
&&{\partial _t}{{\rm{\bar X}}_{ \Downarrow  \Uparrow }} =  - (i{\Delta _{\rm{k}}} + \frac{{{\Gamma _{\rm{k}}}}}{2}){{\rm{\bar X}}_{ \Downarrow  \Uparrow }} - i J {{\rm{\bar X}}_{ \downarrow  \uparrow }} + {\bar F_{ \Downarrow  \Uparrow }}.
\label{eq-1}
\end{eqnarray}
with ${{\rm{\bar X}}_{ \downarrow \downarrow}} = {{\rm{X}}_{ \downarrow \downarrow}}/\sqrt {{p_a}{n_a}}$, ${{\rm{\bar X}}_{ \downarrow {\rm{p}}}} = {{\rm{X}}_{ \downarrow {\rm{p}}}}/\sqrt {{p_a}{n_a}}$, ${\rm{ }}{{\rm{\bar X}}_{ \downarrow  \uparrow }} = {{\rm{X}}_{ \downarrow  \uparrow }}/\sqrt {{p_a}{n_a}}$, ${{\rm{\bar X}}_{ \Downarrow  \Uparrow }} = {{\rm{X}}_{ \Downarrow  \Uparrow }}/\sqrt {{p_b}{n_b}}$, ${{\rm{\bar F}}_{ \downarrow {\rm{p}}}} = {{\rm{F}}_{ \downarrow {\rm{p}}}}/\sqrt {{p_a}{n_a}}$, ${{\rm{\bar F}}_{ \downarrow  \uparrow }} = {{\rm{F}}_{ \downarrow  \uparrow }}/\sqrt {{p_a}{n_a}}$, and ${{\rm{\bar F}}_{ \Downarrow  \Uparrow }} = {{\rm{F}}_{ \Downarrow  \Uparrow }}/\sqrt {{p_b}{n_b}}$.  
The exchange rate $J=\zeta \sqrt{{p_a p_b}{n_a n_b}/4}$ represents the coherent spin-exchange strength between the two atomic gases, which has been experimentally demonstrated to enter the strong-coupling regime~\cite{Shaham2021_Strong}, i.e., $J\gg \Gamma_s$. The modified frequency detunings are given by 
\begin{equation}
{\Delta _{\rm{e}}} = {\tilde\omega_e} - {{\rm{\omega }}_{\rm{p}}},   
{\Delta _{\rm{s}}} = {\tilde\omega_s} - ({{\rm{\omega }}_{\rm{p}}} - {{\rm{\omega }}_{\rm{c}}}),
{\Delta _{\rm{k}}} = {\tilde\omega_k} - ({{\rm{\omega }}_{\rm{p}}} - {{\rm{\omega }}_{\rm{c}}}).
\label{eq-S21}
\end{equation}
Here $\tilde\omega_e$,  $\tilde\omega_s$, and $\tilde\omega_k$ denote the overall frequencies of the energy levels, containing the frequency shifts due to the magnetic field and the spin-exchange interactions.  

The overall frequency difference between the alkali and noble-gas spins can be compensated by a specific magnetic field $B=B_{0}$, resulting in $\tilde\omega_s=\tilde\omega_k$, i.e., $\Delta _{s}=\Delta _{k}$. 
Furthermore, for the convenience of the theoretical analysis, we assume that the detunings satisfy a simple relationship $\Delta _{e}=\Delta _{s}=\Delta _{k}$ at $B=B_{0}$, which can be realized by adjusting the frequency of the control field. Accordingly, we deduce that ${\Delta _{s}}={\Delta _{e}}+{\gamma_{s}}{\widetilde{B}}$, ${\Delta _{k}}={\Delta _{e}}+{\gamma_{k}}{\widetilde{B}}$ hold for a general magnetic field $B=B_{0}+\widetilde{B}$ and tunable probe frequency $\omega_p$.

\section{Reduction of complex susceptibility}

The complex susceptibility is given by
\begin{eqnarray}
\chi \propto \frac{\left\langle{\bar X}_{ \downarrow p} \right\rangle}{\Omega_p} ,
\label{eq-S9}
\end{eqnarray}
By solving the motion equation in the main text, the steady-state solution for the expected value of $\bar X_{ \downarrow {\rm{p}}}$ can be obtained
\begin{eqnarray}
\left\langle{\bar X}_{ \downarrow p} \right\rangle  = \frac{{ - i\Omega_p( { {J}^2 + ( {\frac{{{\Gamma _s}}}{2} + i{\Delta _s}} )( {\frac{{{\Gamma _k}}}{2} + i{\Delta _k}} )} )}}{{{J}^2( {\frac{{{\Gamma _e}}}{2} + i{\Delta _e}} ) + ( {\frac{{{\Gamma _k}}}{2} + i{\Delta _k}} )( {( {\frac{{{\Gamma _e}}}{2} + i{\Delta _e}} )( {\frac{{{\Gamma _s}}}{2} + i{\Delta _s}} ) + {{\left| \Omega  \right|}^2}} )}},    \label{eq-S10}
\end{eqnarray}
where ${\Delta _s} = {\Delta _e} +  \gamma_s\widetilde{B},{\Delta _k} = {\Delta _e} +  \gamma_k\widetilde{B}$.  
In the absence of the coherent spin-exchange interaction between the alkali and noble-gas spins, that is, $J=0$, the complex susceptibility can be reduced to
\begin{eqnarray}
\chi_{J = 0} = - i\eta \frac{{  {\frac{{{\Gamma _s}}}{2} + i{\Delta _s}} }}{{( {\frac{{{\Gamma _e}}}{2} + i{\Delta _e}} )( {\frac{{{\Gamma _s}}}{2} + i{\Delta _s}} ) + {{\left| \Omega  \right|}^2}}} ,
\label{eq-S11}
\end{eqnarray}
which is the susceptibility of the generic EIT configuration, with the constant normalization coefficient $\eta$. 
Through Laplace transform with the variable $\Delta _e$ which can be performed using Mathematica software, we can obtain the transform function of $\chi_{J = 0}$ as
\begin{eqnarray}
\widetilde \chi_{J=0}  = \widetilde \chi_{1,J=0} + \widetilde \chi_{2,J=0},    \label{eq-S12}
\end{eqnarray}
where
\begin{subequations}
\begin{equation}
	\widetilde \chi_{1,J=0} =  - i\eta \frac{{ {\lambda  - 2i \gamma_s\widetilde{B} + {\Gamma _e} - {\Gamma _s}} }}{{2\lambda }}{{\rm{e}}^{ - \frac{1}{4}t( {\lambda  + 2i \gamma_s\widetilde{B} + {\Gamma _e} + {\Gamma _s}} )}} ,  
	\label{eq-S13a}
\end{equation}
\begin{equation}
	\widetilde \chi_{2,J=0} =  - i\eta \frac{{ {\lambda  + 2i \gamma_s\widetilde{B} - {\Gamma _e} + {\Gamma _s}} }}{{2\lambda }}{{\rm{e}}^{\frac{1}{4}t( {\lambda  - 2i \gamma_s\widetilde{B} - {\Gamma _e} - {\Gamma _s}} )}},   
	\label{eq-S13b}
\end{equation}
\end{subequations}
with
\begin{eqnarray}
\lambda  = \sqrt {{{( {{\Gamma _e} - {\Gamma _s} - 2 i  \gamma_s\widetilde{B}} )}^2} - 16{{\left| \Omega  \right|}^2}} .
\label{eq-S14}
\end{eqnarray}
Then, performing inverse Laplace transform on $\widetilde \chi_{1,J=0}$ and $\widetilde \chi_{2,J=0}$, respectively, we can obtain
\begin{subequations}
\begin{equation}
	\chi_{1,J=0} =  - i\eta \frac{{1}}{{i( {{\Delta _e} - \frac{{4 \gamma_s\widetilde{B}{{\left| \Omega  \right|}^2}}}{{{\Gamma _e}^2}}} ) + \frac{{{\widetilde{\Gamma} _e}}}{2}}} ,
	\label{eq-S15a}
\end{equation}
\begin{equation}
	\chi_{2,J=0} =  - \eta \frac{{8( {2 \gamma_s\widetilde{B} - i\frac{{{\Gamma _e}}}{2}} ){{\left| \Omega  \right|}^2}}}{{( {i( {{\Delta _e} +  \gamma_s\widetilde{B}} ) + \frac{{{\widetilde{\Gamma} _s}}}{2}} ){\Gamma _e}^3}},   
	\label{eq-S15b}
\end{equation}
\end{subequations}
where
\begin{subequations}
\begin{equation}
	{\widetilde{\Gamma} _e} = {\Gamma _e} - \frac{{4{{\left| \Omega  \right|}^2}}}{{{\Gamma _e}}},  
	\label{eq-S16a}
\end{equation}
\begin{equation}
	{\widetilde{\Gamma} _s} = {\Gamma _s} + \frac{{4{{\left| \Omega  \right|}^2}}}{{{\Gamma _e}}}. 
	\label{eq-S16b}
\end{equation}
\end{subequations}
Therefore, the susceptibility of EIT configuration can be divided into two terms, both of which are Lorentzian curves with linewidths ${\widetilde{\Gamma} _e}$ and ${\widetilde{\Gamma} _s}$, respectively. The first term represents the general absorption of an optical transition, and the second term corresponds to EIT window.

In the presence of the coherent spin-exchange interaction between the alkali and noble-gas spins, the complex susceptibility can be decomposed into three terms following the above method applied in the case of $J=0$,  that is,
\begin{eqnarray}
\chi  = {\chi _1} + {\chi _2} + {\chi _3}.     \label{eq-S17}
\end{eqnarray}
Here the first two terms correspond to the susceptibility of EIT. And, in the parameter range of interest, the influence of the coherent spin-exchange interaction on these terms is negligible, so we can obtain
\begin{subequations}
\begin{equation}
	\chi_1 =  - i\eta \frac{{1}}{{i( {{\Delta_e} - \frac{{4 \gamma_s\widetilde{B}{{\left| \Omega  \right|}^2}}}{{{\Gamma _e}^2}}} ) + \frac{{{\widetilde{\Gamma} _e}}}{2}}}  , 
	\label{eq-S18a}
\end{equation}
\begin{equation}
	\chi_2 =  - \eta \frac{{8( {2 \gamma_s\widetilde{B} - i\frac{{{\Gamma _e}}}{2}} ){{\left| \Omega  \right|}^2}}}{{( {i( {{\Delta_e} +  \gamma_s\widetilde{B}} ) + \frac{{{\widetilde{\Gamma} _s}}}{2}} ){\Gamma _e}^3}}.  
	\label{eq-S18b}
\end{equation}
\end{subequations}
The third term represents the signal induced by the coherent spin-exchange interaction, and in the corresponding frequency region, the sum of the first two terms can be considered to remain a constant value at $\Delta _e=0$. Therefore, in the derivation of $\chi_3$, we can substitute the conditions $\Delta _e=0, \Delta _s=\gamma_s\widetilde{B} $ into the Laplace transform, and obtain
\begin{eqnarray}
\chi_3 =  - i\eta \frac{{4{J}^2{{\left| \Omega  \right|}^2}}}{{( {i( {{\Delta _e} - \bar{\omega} } ) + \frac{{{\widetilde{\Gamma} _k}}}{2}} ){{( {i \gamma_s\widetilde{B}{\Gamma _e} + \frac{1}{2}{\Gamma _e}{\Gamma _s} + 2{{\left| \Omega  \right|}^2}} )}^2}}},
\label{eq-S19}
\end{eqnarray}
where 
\begin{subequations}
\begin{equation}
	\bar{\omega}  = -{\gamma_k}\widetilde{B} + {\gamma_s}\widetilde{B} \frac{{{J}^2{\Gamma _e}^2}}{{{ \gamma_s^2\widetilde{B}^2 }{\Gamma _e}^2 + {{( {\frac{1}{2}{\Gamma _e}{\Gamma _s} + 2{{\left| \Omega  \right|}^2}} )}^2}}},   
	\label{eq-S20a}
\end{equation}
\begin{equation}
	{\widetilde{\Gamma} _k} = {\Gamma _k} + \frac{{{J}^2{\Gamma _e}( {{\Gamma _e}{\Gamma _s} + 4{{\left| \Omega  \right|}^2}} )}}{{{\gamma_s^2\widetilde{B}^2}{\Gamma _e}^2 + {{( {\frac{1}{2}{\Gamma _e}{\Gamma _s} + 2{{\left| \Omega  \right|}^2}} )}^2}}}
	\label{eq-S20b}
\end{equation}
\end{subequations}
are the frequency shift and width of $\chi_3$, respectively.

\section{NSIT in a wide parameter range}

\begin{figure}
\includegraphics[draft=false, width=0.95\columnwidth]{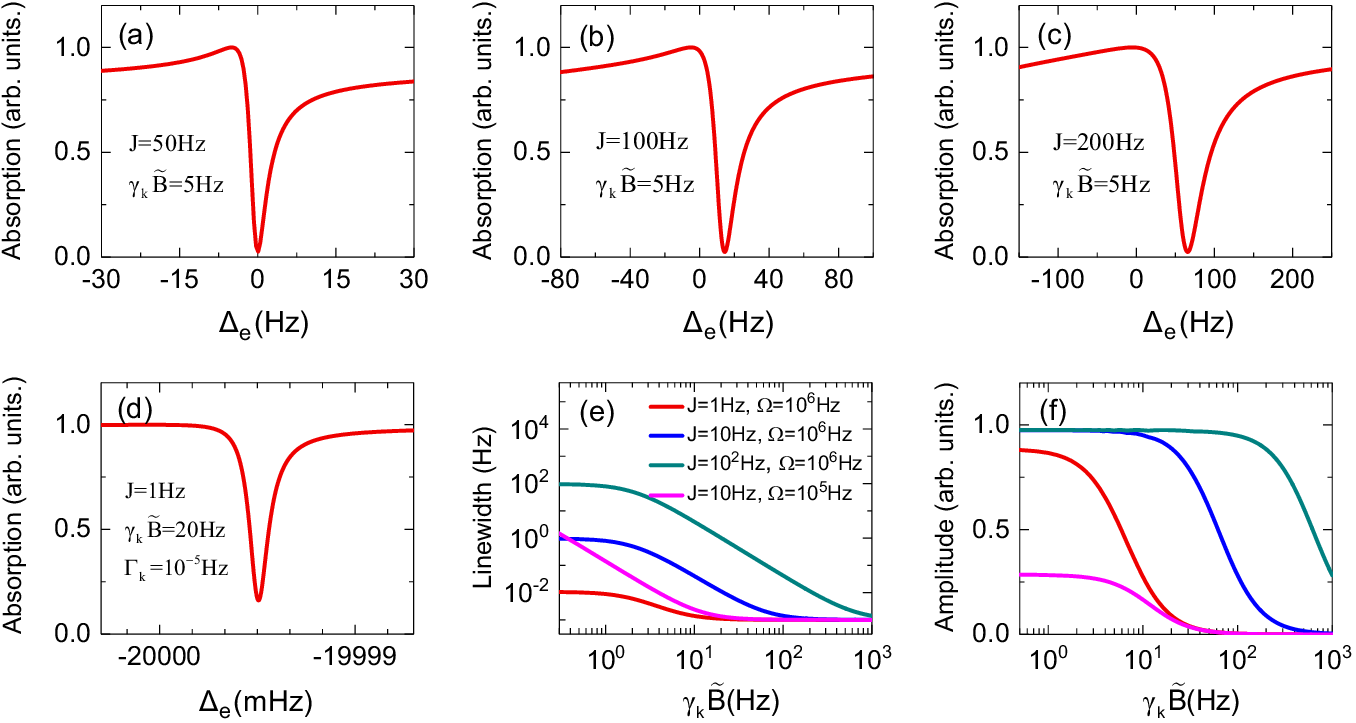}
\caption{(Color online)  (a)-(d) Normalized absorption spectrums of the optical
transition $\left\vert \downarrow \right\rangle \leftrightarrow \left\vert
p\right\rangle $ for different $J$ and $\widetilde{B}$, where NSIT signals are revealed.
(f) Linewidth and (g) amplitude of NSIT signals as functions of $\widetilde{B}$ for different $J$ and $\Omega$. 
Other parameters are the same as Fig.~4 of the main text. 
\label{figS1}}
\end{figure}

We now consider the absorption property of the optical transition $\left|  \downarrow  \right\rangle  \leftrightarrow \left| p \right\rangle$ over a wide parameter range.
According to Eq.~(\ref{eq-S20b}), one can deduce that $J$ and $\widetilde{B}$ have significant effects on the width and amplitude of the NSIT signal, which can also be visualized from Fig.~\ref{figS1}, and then on the slopes of the dispersion curves (not shown here).
Specifically, one can see from Figs.~\ref{figS1} (a)-(e) that by varying $J$, the linewidth can vary from sub-mHz to tens of Hz or more.  
By adjusting $J$ to an appropriate intensity as shown in Figs.~\ref{figS1}(b) and (c), the signal width can reach the order of tens of Hz or even greater, i.e., the width range of EIT window~\cite{Brandt1997_Buffer, Erhard2001_Buffer}, and simultaneously the signal amplitude is well maintained, which indicates that NSIT signal can become as pronounced as EIT signal in atomic spectrums.
Meanwhile, according to the discussion in the main text, we know that when $J$ is small, the signal width can approach the minimum value, i.e., the decoherence rate $\Gamma _k$ of noble-gas spin. 

The EIT window is a direct manifestation of the two-photon resonance region, and NSIT is a signal induced by the two-photon transition together with the spin-exchange interaction. Therefore, the separation between the EIT and NSIT signals $(\gamma_s-\gamma_k)\widetilde B \approx \gamma_s \widetilde B$ has a significant influence on the amplitude of the NSIT signal.
When the separation is much larger than the width of the EIT window, i.e., $\gamma_s \widetilde B\gg \Omega^2/\Gamma_e $ (see the analytic expression of $\chi _3$ in Eq.~(\ref{eq-S19})), the signal amplitude reduces significantly, along with the signal width, as shown in Figs.~\ref{figS1}(e) and (f). 
Accordingly, to obtain a strong NSIT signal, it is necessary to make sure that the separation between the EIT and NSIT signals is not too large. 
Meanwhile, one can see from Figs.~\ref{figS1}(a)-(d) that the amplitude of the NSIT signal can be well maintained when the spin-exchange rate $J$ and the magnetic field $\widetilde{B}$ vary in a wide range.

\section{Doppler effect}

\begin{figure}
	\includegraphics[draft=false, width=0.95\columnwidth]{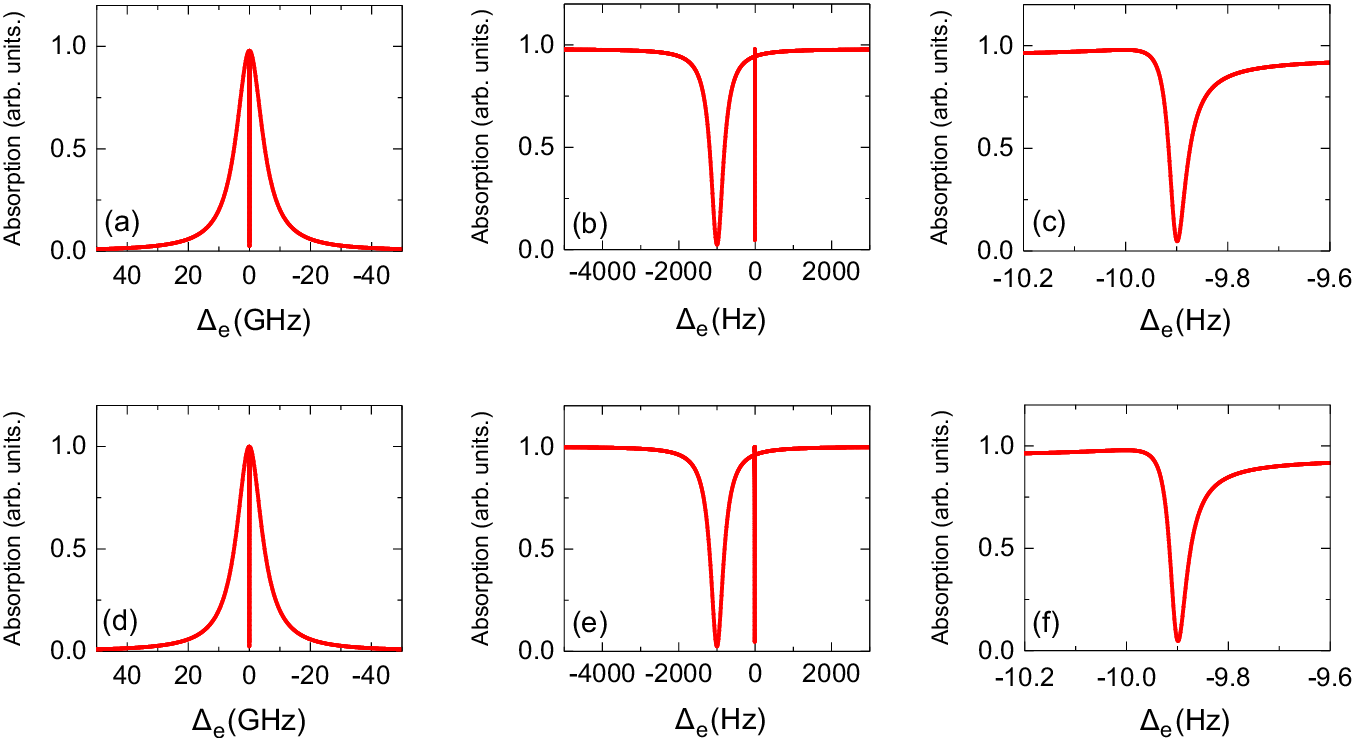}
	\caption{(Color online) Normalized absorption spectrum of the optical
transition $\left| \downarrow \right\rangle \leftrightarrow \left| p
\right\rangle$ under Doppler effect, where the broad single-photon resonance signal, EIT signal, and NSIT signal are exhibited in different frequency scales as shown in (a), (b), and (c), respectively.  
The absorption spectrums without considering Doppler effect are shown in (d), (e), and (f), corresponding to (d), (e), and (f), respectively.
The parameters are the same as Fig. 4(c) of the main text.
		\label{figS2}}
\end{figure}

According to the previous discussion, the atomic absorption spectrum contains three spectral components, i.e., the broad absorption peak, the EIT window, and the NSIT/NSIA signal caused by the long-lived noble-gas nuclear spins. These three spectral components are respectively governed by the three frequency detunings $\Delta _{\rm{e}}, \Delta _{\rm{s}}$, and $\Delta _{\rm{k}}$ in Eq.~(\ref{eq-S21}). Due to the Doppler effect, an atom moving towards the light field (with frequency $\omega_L$, propagating along the z-axis) with velocity $v_z$ is affected by the frequency detuning upshifted to $\omega_L \pm\omega_L v_z/c$, where the plus or minus sign depends on whether the atom travels in the reverse or same direction as the light propagation~\cite{Li1995_Observation, Gea-Banacloche1995_Electromagnetically}. For a $\Lambda$-shape system, the two-photon Doppler-free technique requires that the control field and the probe field propagate in the same direction. An atom moving towards the light propagation with velocity $v_z$ is affected by the probe-field frequency upshifted to $\omega_p \pm\omega_p v_z/c$ and control-field frequency upshifted to $\omega_p \pm\omega_p v_z/c$.  Therefore, considering the Doppler shift, the three frequency detunings in Eq.~(\ref{eq-S21}) become 
\begin{equation}
\Delta_{e} = {{\rm{\tilde\omega_e }}} - ({{\rm{\omega }}_{\rm{p}}} + \frac{{{v_z}}}{c}{{\rm{\omega }}_{\rm{p}}}), \\
{\Delta _{\rm{s}}} = {{\rm{\tilde\omega_s }}} - ({{\rm{\omega }}_{\rm{p}}} - {{\rm{\omega }}_{\rm{c}}} + \frac{{{v_z}}}{c}({{\rm{\omega }}_{\rm{p}}} - {{\rm{\omega }}_{\rm{c}}})),  \\
{\Delta _{\rm{k}}} = {{\rm{\tilde\omega_k }}} - ({{\rm{\omega }}_{\rm{p}}} - {{\rm{\omega }}_{\rm{c}}} + \frac{{{v_z}}}{c}({{\rm{\omega }}_{\rm{p}}} - {{\rm{\omega }}_{\rm{c}}})).
\label{eq-S22}
\end{equation}

Because of the Doppler shift, the complex susceptibility becomes 
\begin{equation}
{\chi_D} = \int_{ - \infty }^{ + \infty } \chi(v_z)f({{v_z}})d{v_z},
\label{eq-S23}
\end{equation}
In a vapor cell, atomic velocities will be distributed according to the Maxwell–Boltzmann distribution, which takes the form  $f\left( {{v_z}} \right) = \frac{c}{\mu }{e^{ - {v_z^2}/{\mu ^2}}}$. The most probable velocity $\mu$ satisfies ${\rm{\mu }} = \sqrt {2{k_B}T/m}$, where $m$ is the atomic mass, $k_B$ is the Boltzmann constant, and $T$ is the temperature. Substituting Eqs.~(\ref{eq-S9}) and (\ref{eq-S22}) into Eq.~(\ref{eq-S23}), one can obtain the complex susceptibility under the Doppler effect. For visualization, we show the absorption spectra of alkali-metal atoms under the Doppler-free configuration in Fig.~\ref{figS2}. Of these, the EIT and NSIT signals are close-up in Figs.~\ref{figS2}(b) and (c), respectively. In contrast, the corresponding spectral patterns without considering the Doppler effect are also shown in Figs.~\ref{figS2}(d), (e), and (f), respectively. 

From Eq.~(\ref{eq-S22}), one can see that in the Doppler-free configuration where the frequencies of the control and probe light fields are degenerate, the Doppler shift is canceled in both two-photon detunings $\Delta_s$ and $\Delta_k$ that correspond to EIT and NSIT signals, respectively, and is only retained in single-photon detuning $\Delta_e$ that corresponds to the broad single-photon resonance signal. Further, from the expression for the complex susceptibility $\chi$ in Eqs.~(\ref{eq-S9}) and (\ref{eq-S10}), one can see that the factor associated with $\Delta_e$ is $(\Gamma_e/2+i \Delta_e)$. 
Thereby, the influence of the Doppler effect on the complex susceptibility is reflected in the single-photon transition $\left| \downarrow \right\rangle \leftrightarrow \left| p \right\rangle$ and thus determined by the ratio of the single-photon Doppler width $\Gamma_D$ to the total relaxation rate $\Gamma_e$ of the excited state, with ${\Gamma _{\rm{D}}} = {\omega _p}\sqrt {\frac{{8 \text{Ln[2]}{k_B}T}}{{m{c^2}}}}$.

Due to the pressure broadening, the relaxation rate of the alkali-metal excited state reaches $10^{10}$ Hz, which significantly exceeds the single-photon Doppler broadening of about $10^9$ Hz (e.g., for ${}^{87}\rm{Rb}$ vapor at 500K). Consequently, the Doppler shift in $(\Gamma_e/2+i \Delta_e)$ is unresolved, and thus the complex susceptibility and the corresponding atomic spectral pattern remain essentially unchanged after performing the integral over the velocity in Eq.~(\ref{eq-S23}), as shown in Figs.~\ref{figS2}(a) and (d).   
As for the EIT window due to the two-photon resonance effect, since the Doppler shift is canceled, one can see that the EIT signal including the Doppler effect shown in Fig.~\ref{figS2}(b) does not exhibit significant broadening comparing the spectral patterns excluding the Doppler effect shown in Fig.~\ref{figS2}(e) and is well maintained. Therefore, the EIT signal can be well protected when adopting the Doppler-free configuration of the laser fields.
Similar to the EIT effect, NSIT is also a two-photon resonance effect, originating from the interaction of the light field with the noble-gas nuclear spin mediated via the alkali-metal atom. This ensures that the Doppler-free technique remains valid for this effect, and thus, the NSIT window is well maintained, as shown in Fig.~\ref{figS2}(c).

\section{Imperfect polarization}

\begin{figure}
	\includegraphics[draft=false, width=0.56\columnwidth]{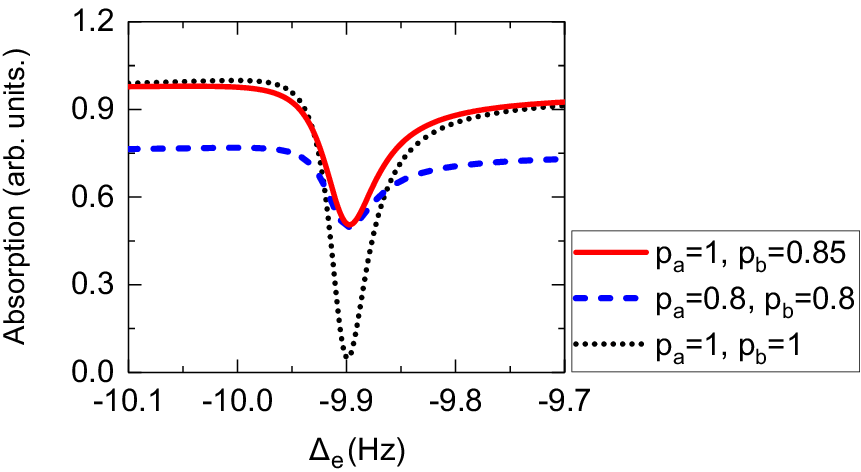}
	\caption{(Color online)  Normalized absorption spectrum of the optical transition $\left| \downarrow \right\rangle \leftrightarrow \left| p \right\rangle$ when the alkali-metal and noble-gas spins have various polarization degrees. The spectral pattern of the NSIT signal for perfect polarization, i.e., Fig. 4(c) of the main text, is shown as a comparison. 
		\label{figS3}}
\end{figure}

While we previously focused on the case of two atomic ensembles being perfectly polarized (i.e., ${p_{\rm{a}}} \to 1, {p_{\rm{b}}} \to 1$), imperfect polarization can have an impact on the NSIT effect. 
In Fig.~\ref{figS3}, we show the NSIT signal when the alkali-metal and noble-gas spins are imperfectly polarized (i.e., ${p_{\rm{a}}} < 1, {p_{\rm{b}}} < 1$), where the imperfect polarization can be simulated by introducing incoherent population transfer between different spin states. It reveals that the NSIT signal remains under imperfect polarization, although the amplitude of the signal decreases to some extent. This phenomenon is understandable based on the fact that NSIT effect is based on the quantum interference consisting of the two-photon transition and the coherent spin-exchange interaction. When a part of the atomic population is incoherently distributed in different spin states due to imperfect polarization, the atomic population involved in the coherent physical process corresponding to NSIT effect inevitably decreases. As a result, the amplitude of the NSIT signal decreases. High polarization degrees of alkali and noble-gas spins (e.g., ${p_{\rm{a}}} \to 1, {p_{\rm{b}}} = 0.85$) have been experimentally demonstrated~\cite{Dideriksen2021_Room, Chen2014_On}, which enables a clear NSIT signal to be accessible.

\bibliography{bib_supple}